\documentclass[%preprint, linenumbers,
superscriptaddress,
 amsmath,amssymb,
 aps, physrev,
twocolumn]{revtex4-2}

\usepackage{xcolor}
\usepackage{graphicx}% Include figure files
\usepackage{dcolumn}% Align table columns on decimal point
\usepackage{bm}% bold math
\usepackage{graphicx}             
\usepackage[caption=false]{subfig}

\begin{document}

%\preprint{APS/123-QED}

\title{\textbf{Polaron transport and Verwey transition in magnetite}}

\author{Nikita Fominykh}
 \email{fominykh.na@phystech.edu}
\affiliation{Joint Institute for High Temperatures of Russian Academy of Sciences, Izhorskaya st. 13 bldg. 2, Moscow, 125412, Russia}
\affiliation{Moscow Institute of Physics and Technology, Institutsky lane 9, Dolgoprudny,
141700, Russia} 
\author{Vladimir Stegailov}%
 \email{stegailov@jiht.ru}
\affiliation{Joint Institute for High Temperatures of Russian Academy of Sciences, Izhorskaya st. 13 bldg. 2, Moscow, 125412, Russia}
\affiliation{Moscow Institute of Physics and Technology, Institutsky lane 9, Dolgoprudny,
141700, Russia} 
\affiliation{HSE University, Myasnitskaya Ulitsa 20, Moscow 101000, Russia}

\date{\today}% It is always \today, today,

\begin{abstract}
The enigmatic puzzle of the Verwey transition in magnetite Fe$_3$O$_4$ has been unresolved for almost a century. We present an ab initio-based model of the polaron transport combining kinetic Monte Carlo and molecular dynamics calculations to directly describe the coupling of polarons with lattice vibrations. Contrary to the Ihle-Lorentz small-polaron model, we find no significant change in the band structure across the Verwey transition, however, trimeron hopping is observed. The proposed model provides dc-conductivity in agreement with experimental data across the Verwey transition.
\end{abstract}

\maketitle

%\tableofcontents
The Verwey transition in magnetite Fe$_3$O$_4$ has remained an enigmatic puzzle for almost a century. The complex lattice distortion from the cubic Fd$\bar{3}$m to the monoclinic Cc lattice at a temperature below $T_V \sim 120~\text{K}$ is accompanied by a decrease in electrical conductivity by a factor of 100 after cooling~\cite{verwey1939electronic,verwey1941electronic,senn2012charge}.
Enormous research efforts have been devoted to experimental and theoretical studies of this transition~\cite{walz2002verwey,garcia2004verwey}. However, many physical properties such as lattice structure, band gap, and charge ordering are subjects of longstanding debate. One of the most emblematic controversies is that magnetite served as an initial prototype material for the Mott concept of the metal-insulator transition~\cite{mott1968metal}, although modern experiments manifest this transition as a semiconductor-semiconductor type~\cite{schrupp2005high,pimenov2005terahertz,prozorov2023response}. At the same time, metallic interpretations, which could be more relevant to pressure-induced transition~\cite{mori2002metallization,rozenberg2006origin,kozlenko2019magnetic}, are used even in recent studies~\cite{taguchi2015temperature,kukreja2018orbital, tiwari2026near, du2026synthesis}, highlighting the need for further studies.

The phenomenological Ihle-Lorentz microscopic theory~\cite{ihle1986small} of small-polaron conductivity in magnetite introduce conductivity picture before modern ab initio data were available. It combines small-polaron hopping and band conduction taking into account short-range ordering due to polaron interactions, which results in the appearance of an additional polaron sub-band
above the Verwey transition. For quite some time, the question of charge ordering remained unsettled even in the case of the low-temperature phase.
The later development of ab initio approaches for treating strong electronic correlations~\cite{anisimov1991band} allowed significant progress in magnetite charge-orbital models~\cite{leonov2004charge,jeng2004charge,leonov2006electronic,zhou2010first}, which were in agreement with the Kugel-Khomskii model~\cite{kugel1982jahn,igoshev2023multiorbital}.
However, the small charge disproportion of the order of $0.1e$ between the Fe$^{3+}$ and Fe$^{2+}$ formal valence states and small structural distortions of the order of $0.1$~{\AA} together with sample defects hindered experimental research and correct low-T structure was obtained later~\cite{senn2012charge,senn2012electronic}.
A characteristic feature of the low-T structure is a complex network of linear orbital molecules composed of formal Fe$^{3+}$-Fe$^{2+}$-Fe$^{3+}$ units called trimerons~\cite{attfield2022magnetism}.
Moreover, short-range structural fluctuations, which could be linked to the presence of trimerons, were reported at temperatures much higher than $T_V$ from diffuse scattering~\cite{bosak2014short}, x-ray scattering~\cite{huang2017jahn,perversi2019co,elnaggar2020possible}, ultrafast electron diffraction~\cite{wang2023verwey}, and scanning near-field optical microscopy~\cite{tiwari2026near}.
The interplay of electronic and lattice degrees of freedom is the central topic for unveiling the mechanism of the Verwey transition. In that vein, polaronic band~\cite{ihle1986small}, electron-phonon coupling~\cite{piekarz2006me,rowan2009hybrid,hoesch2013anharmonicity}, the order-disorder transition~\cite{borroni2017mapping}, soft electronic modes~\cite{baldini2020discovery}, trimeron-phonon coupling~\cite{piekarz2021trimeron}, and electronic nematicity~\cite{wang2023verwey} are proposed to be crucial for the Verwey transition.

In this Letter, we would like to take a novel look at the polaron transport and the Verwey transition via the DFT+U based kinetic Monte Carlo (kMC) and molecular dynamics (MD) calculations, extending our previous work on the low-T phase static calculations of magnetite~\cite{fominykh2025trimeron}. 
The combination of these two methods provides a unique opportunity to study polaron transport at the ab initio level both below and above the Verwey transition. The MD approach allows us to model directly the coupling of polarons with lattice vibrations at finite temperatures, which, to the best of our knowledge, has not been applied to this problem before.

Spin-polarized DFT+U calculations are performed using the projector-augmented wave method~\cite{blochl1994projector,kresse1999ultrasoft} implemented in VASP~\cite{kresse1993ab,kresse1996efficient}. The cutoff energy of the plane-wave basis is 550 eV. The exchange-correlation energy is used in the generalized gradient approximation~\cite{perdew1996generalized}. Strong correlations are taken into account with the Hubbard $U_\text{eff}=U-J$ correction applied to the 3d electrons of Fe based on the rotation-invariant Dudarev approach~\cite{dudarev1998electron}. Based on previous work, we use $U_\text{eff}=3.8~\text{eV}$~\cite{fominykh2025trimeron,fominykh2025influence,fominykh2023polarons}. Calculations are performed for the shape-relaxed $\sqrt{2}a \times \sqrt{2}a \times 2a$ $Cc$ supercell with 112 atoms, where $a$ is the undistorted cubic cell lattice parameter. Static polaron calculations use $\Gamma$-centered $2\times2\times2$ k-mesh, while MD calculations use $\Gamma$-point only.

\begin{figure}
    \centering
    \includegraphics[width=0.95\linewidth]{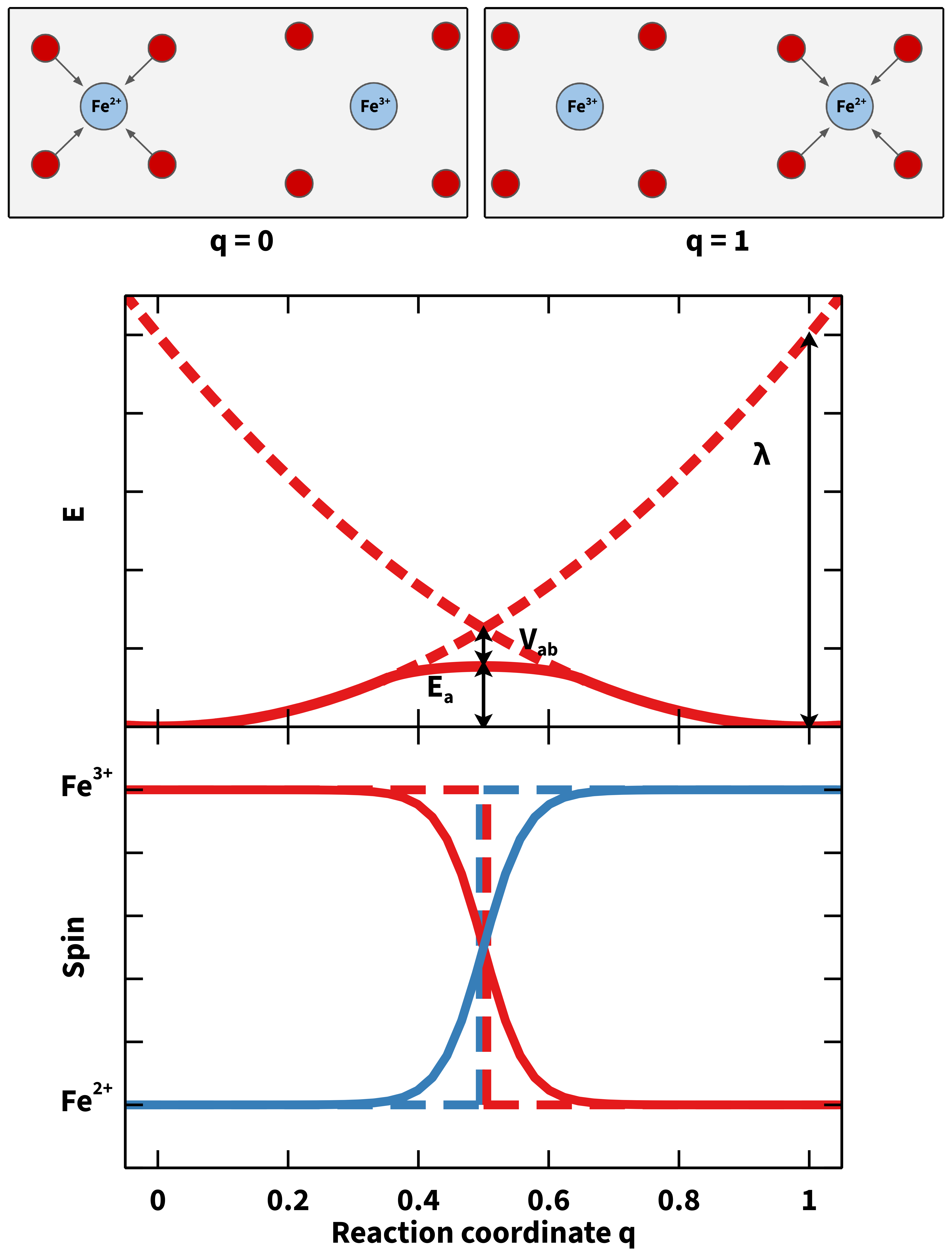}
    \caption{The simplified principal scheme of polaron hopping. Dashed lines represent diabatic hopping, solid lines represent adiabatic hopping. Top panel shows energy profile scheme and simplified picture of polaron distortion for $q=0$ and $q=1$. Activation energy $E_a$, reorganization energy $\lambda$ and electronic coupling $V_{ab}$ are shown by arrows. Bottom panel shows spin values for initial and final states. }
    \label{fig:scheme}  
\end{figure}

First, we precisely investigate polaron hopping in the low-T phase. We treat the small polaron in magnetite as a self-trapped excess electron or hole charge. The $Cc$ structure has in the octahedral sublattice 16 Fe$^{3+}$ sites available for electron polaron formation and similarly 16 Fe$^{2+}$ sites available for hole polaron formation. Considering only nearest-neighbor hopping, there are 22 elementary paths for both hole and electron polaron hopping. For each of them, excess charge is localized at the corresponding sites, after which the energy surfaces are calculated along linearly interpolated coordinates between the localized states $\mathbf{R_{q}} = (1-q)\mathbf{R_{q=0}} + q\mathbf{R_{q=1}}$. For all polarons, diabatic energy surfaces were obtained (Fig.S1 in SM). The simplified principal scheme is shown in Figure~\ref{fig:scheme}. From these energy surfaces, elementary activation energies for all possible elementary paths are obtained according to the Holstein-Mott theory~\cite{emin1969studies,austin1969polarons,holstein2000studies} and the Marcus approach~\cite{marcus1985electron,marcus1993electron} (see more details in~\cite{fominykh2025trimeron}).

To obtain a macroscopic-like average activation energy, we perform kinetic Monte Carlo (kMC) calculations over the elementary activation energies. To preserve consistency with the following MD calculations, we obtain the activation energy from the Arrhenius dependence of the hopping frequency in the kMC model (details are provided in SM). The obtained activation energies are $E_a^{LT~ep} =  0.15~\text{eV}$ and $E_a^{LT~hp} = 0.18~\text{eV}$ for electron and hole polarons (Fig.S2 in SM), respectively. These activation energies are in good agreement with our previous estimate~\cite{fominykh2025trimeron} and experimental data $E_a^{LT~exp} = 0.10-0.17~\text{eV}$~\cite{matsui1977specific,prozorov2023response}.

\begin{figure*}
   \centering
   \includegraphics[width=0.70\linewidth]{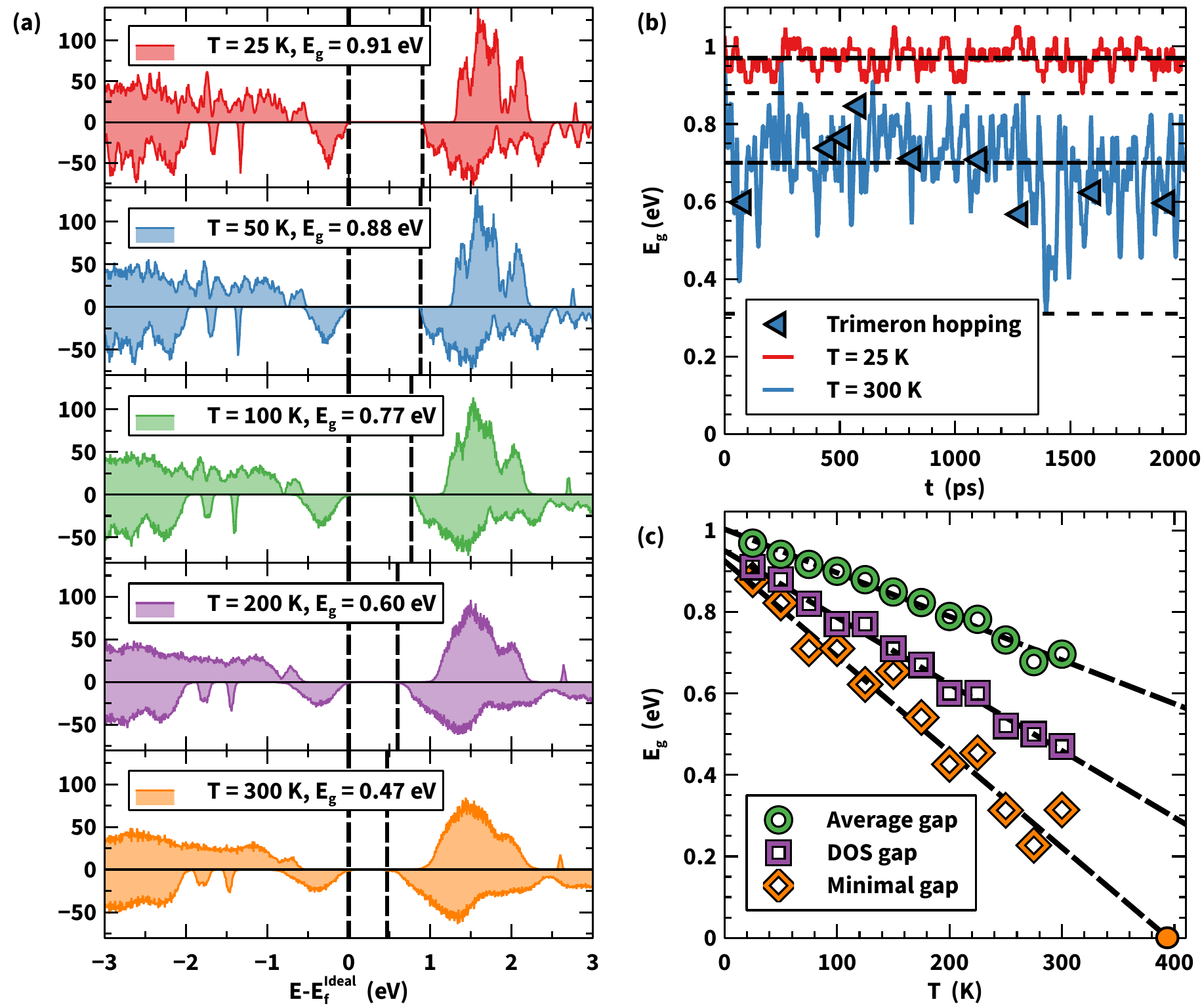}
   \caption{(a) Electronic DOS averaged along each MD trajectory for a given temperature (dashed lines correspond to gap with 1 a.u. threshold), $E_g$ values for the corresponding DOS gap are shown. (b) The example of time dependent instantaneous gap for 25~K and 300~K temperatures. Triangles show trimeron hopping events, dashed lines illustrate an average $E_g$ value, dotted lines show a minimal $E_g$ value. (c) Temperature dependent average $E_g$ values, DOS $E_g$ values and minimal $E_g$ values.}
   \label{fig:dos}  
\end{figure*}

We perform ab initio MD calculations with a 2~fs time step. The constant temperature 2~ps long MD trajectories are calculated for the temperature range of 25-300~K with step 25~K. Nose-Hoover thermostat and the single volume approach are used.
For MD, unlike static calculations, supercells without excess charge are considered.
The temperature-dependent average electronic densities of states (DOS) are shown in Figure~\ref{fig:dos}a. 
The instantaneous highest occupied Kohn-Sham orbital-lowest unoccupied Kohn-Sham orbital gap was obtained (an example is shown in Figure~\ref{fig:dos}b).
The temperature-dependent average $E_g$, minimal $E_g$, and DOS $E_g$ are shown in Figure~\ref{fig:dos}c.
In contrast to the earlier picture of the formation of polaronic bands~\cite{ihle1986small}, it is quite surprising that the DOS changes smoothly with temperature and is mainly affected by a slight thermal broadening. Both average and minimum gaps decrease linearly with temperature without changes around $T_V$. The complex interplay and difficulties in distinguishing between polaronic and band effects were addressed earlier, for example, in~\cite{gasparov2000infrared,schrupp2005high}.

\begin{figure*}
    \centering
    \includegraphics[width=0.93\linewidth]{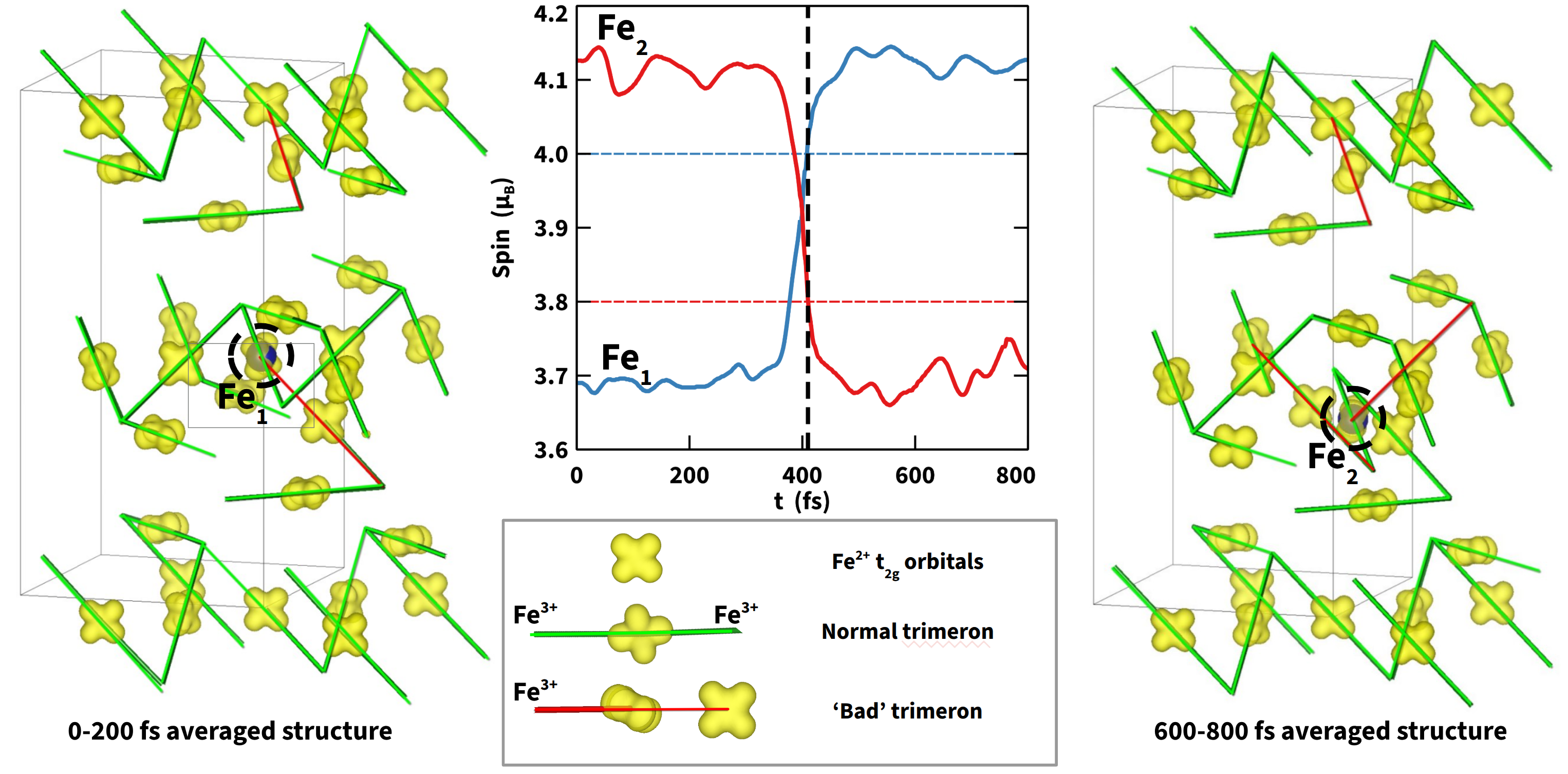}
    \caption{Illustration of one trimeron hopping event. Time dependent spin values are shown for two hopping sites Fe$_1$ and Fe$_2$ on the MD trajectory at $T=150~\text{K}$. The vertical dashed line shows the hopping moment according to the criterion $\Delta\mu>0.3\mu_B$, which is shown by the horizontal dashed lines. The trimeron structure before and after the hopping event based on the averaged atomic positions (with visualized Fe$^{2+}$ isosurfaces of occupied $t2_g$ orbitals). The normal and `bad' trimerons are marked~\cite{senn2012charge,fominykh2025trimeron}. }
    \label{fig:arr}  
\end{figure*}

Despite the fact that the band structure does not show any critical points near $T_V$, along the calculated MD trajectories, thermally activated Fe$^{2+}$/Fe$^{3+}$ orbital hopping is observed in the octahedral sublattice starting from about 150~K. 
We define trimeron hopping as a local change in the formal charge order (Figure~\ref{fig:arr}). Although this process is physically equivalent to the hopping of an intrinsic polaron, we introduce this concept to more clearly distinguish it from the polaron transport of excess charge in the low-T phase without the destruction of the long-range trimeron order. We assume that active local charge reordering could be attributed to the high-T phase, and that, due to active charge reordering, excess charge hopping is indistinguishable from trimeron hopping in terms of polaron hopping properties.

For trimeron hopping, adiabatic spin behavior is observed in contrast to static low-T diabatic polaron hopping energy profiles. There is a continuous spin transition from $\mu_{Fe^{2+}} \sim 3.7~\mu_B$ to $\mu_{Fe^{3+}} \sim 4.1~\mu_B$ for times up to 100 fs. We use $\Delta\mu > 0.3~\mu_B$ as a criterion to determine trimeron hopping events (example of marked events shown in Figure~\ref{fig:dos}b, example of the spin transition shown in Figure~\ref{fig:arr}).
The calculated trimeron hopping frequencies show a pronounced Arrhenius dependence with the activation energy $E_a^{MD} = 0.06~\text{eV}$ (S4 in SM), which is in good agreement with the high-T experimental conductivity activation energy $E_a^{HT~exp} = 0.05-0.07~\text{eV}$~\cite{matsui1977specific,prozorov2023response}. To obtain further data on this mechanism, we have also performed calculations with $U_\text{eff}=3.5~\text{eV}$ and $U_\text{eff}=3.2~\text{eV}$ (see SM). The lowering $U_\text{eff}$ reduces the activation energy and the threshold temperature for the start of the trimeron hopping up to 75~K.

\begin{figure}
    \centering
    \includegraphics[width=0.95\linewidth]{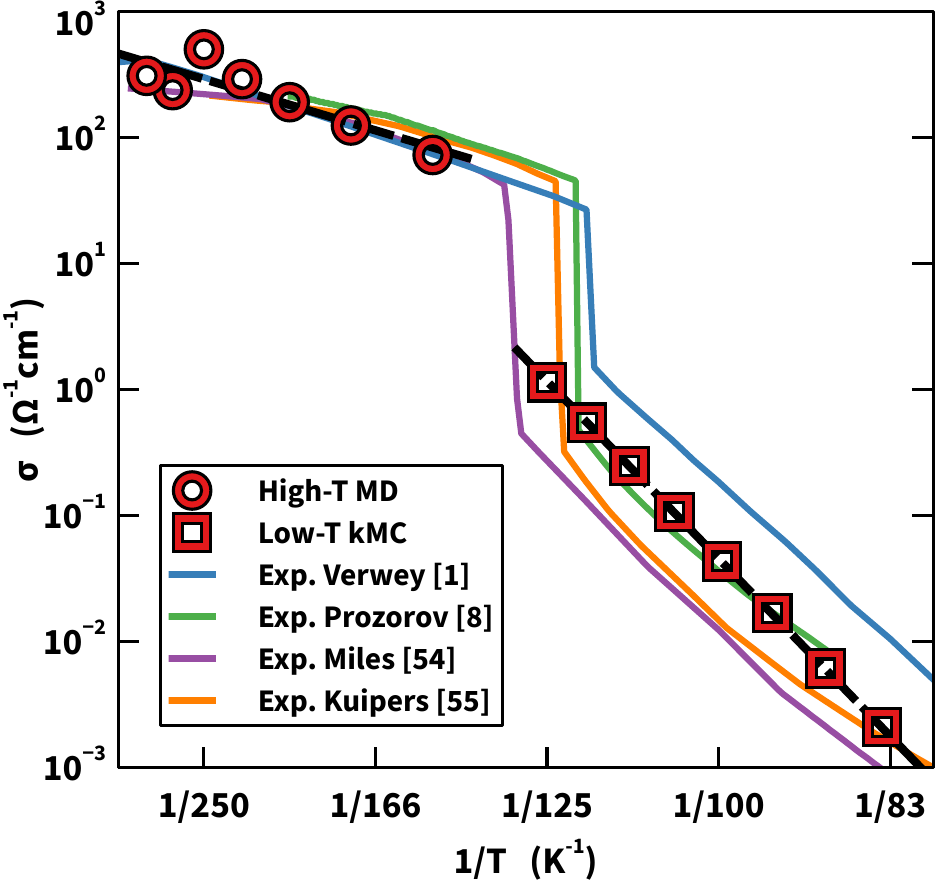}
    \caption{Conductivity obtained using trimeron hopping rates from high-T MD assuming adiabatic hopping, and using low-T kMC hopping rates assuming nonadiabatic hopping.}
    \label{fig:conductivity}  
\end{figure}

The conductivity from hopping data is derived by $\sigma = ne\mu$, $\mu=\frac{d^2k_{hop}}{k_BT}$, where $n$ is the carrier concentration, $d$ is the hopping length, and $k_{hop}$ is the hopping frequency. Assuming nonadiabatic hopping in the low-temperature phase 
\[k^{LT}_{hop} \approx \frac{V^2_{ab}}{\hbar}\sqrt{\frac{\pi}{4k_BTE_a^{LT}}}\exp\left({-\frac{E_a^{LT}}{k_BT}}\right),\] 
where $V_{ab}$ is the electronic coupling~\cite{holstein2000studies,marcus1985electron}. Assuming $V_{ab}\approx E_a^{LT}-E_a^{HT}$, we obtain 
\[\sigma^{LT} = \frac{ned^2}{k_BT}\frac{(E_a^{LT}-E_a^{HT})^2}{\hbar}\sqrt{\frac{\pi}{4k_BTE_a^{LT}}}\exp\left({-\frac{E_a^{LT}}{k_BT}}\right),
\]
\[\sigma^{HT} = \frac{ned^2k_{hop}^{MD}}{k_BT}.\]
Assuming $n = 1$ electron per formula unit and $d = 3${\AA}, we obtain conductivity values in good agreement with the experimental data~\cite{verwey1939electronic,miles1957dielectric,kuipers1979electrical,prozorov2023response} (see Figure~\ref{fig:conductivity}).

The results obtained  could provide a refined interpretation of polaron charge transport around the Verwey transition: below $T_V$, polaron hopping occurs nonadiabatically in the frozen trimeron-ordered $Cc$ structure with $E_a^{LT}=0.15~\text{eV}$, above $T_V$, polaron hopping occurs adiabatically with a significant drop in the activation energy to $E_a^{HT}=0.06~\text{eV}$, which is accompanied by the destruction of the ordered trimeron structure. 
This interpretation refines the earlier concepts~\cite{wang2023verwey,piekarz2021trimeron,borroni2017mapping,hoesch2013anharmonicity,piekarz2006me} linking the Verwey transition and polaron hopping mechanism. 
Another interesting finding, somewhat beyond the scope of the Verwey transition, is the vanishing of the minimal gap near 400~K (Figure~\ref{fig:dos}b). This fact could be connected with temperature-driven self-doping~\cite{elnaggar2021temperature}.

The limitations of the approach used should also be noted. 
%In the DFT+U approach we consider only effective one-electron local static strong correlations. 
Despite the fact that supercells are relatively large, size effects could be quite significant. Unfortunately, it is the upper bound of the achievable computational time for ab initio MD. The same limitation applies to the picosecond trajectory lengths. However, the good agreement with the available experimental data supports our model and results.  
The high-T phase should also be discussed. Although we associate trimeron hopping with the high-T phase, we cannot directly observe a clear structural transition to cubic Fd$\bar{3}$m within the time and supercell scales currently available. However, our approach could be considered in terms of local symmetry-broken motifs that average to cubic symmetry on larger scales (similar cases are reported for a wide range of quantum oxides~\cite{xiong2025symmetry}). Furthermore, this is complementary to the observed short-range structural fluctuations~\cite{bosak2014short,huang2017jahn,perversi2019co,elnaggar2020possible,wang2023verwey,tiwari2026near}.

In summary, to study the Verwey transition, polaron transport in magnetite at the temperature range of $25-300~\text{K}$ has been investigated by ab initio MD and static polaron calculations combined with a kMC model based on the obtained nearest-neighbour polaron hopping activation energies. From the kMC model, the diabatic activation energy $E_a^{LT} = 0.15~\text{eV}$ was obtained. 
The MD calculations show that there is no significant changes in the electron density of states, nevertheless, temperature-activated trimeron hopping occurs and disrupts the ordered low-T trimeron structure at temperatures above 150~K. This process could be attributed to the high-T phase and is accompanied by the activation energy drop to $E_a^{HT}=0.06~\text{eV}$.
The proposed change from nonadiabatic hopping below the Verwey transition to adiabatic hopping above the transition provides, to our knowledge, the first ab initio-based model yielding conductivity values in agreement with experiment on both sides of the transition.
The obtained results emphasize the crucial role of the interplay between electronic and lattice degrees of freedom and indicate a correlation between the disruption of the trimeron ordering and the hopping adiabatization during the Verwey transition.

This work was prepared with the support of the Ministry of Science and Higher Education of the Russian Federation (State Assignment No.~075-00270-24-00) and, in part, within the framework of the HSE University Basic Research Program (optimization of computational efficiency). The work was supported by the Foundation for the Advancement of Theoretical Physics and Mathematics “BASIS” (Grant No.~25-1-5-128-1). 
The authors gratefully acknowledge the access to the resources of the Supercomputer Centre of JIHT RAS and the HPC facilities at HSE University~\cite{kostenetskiy2021hpc}.

%\bibliography{liter} % Produces the bibliography via BibTeX.


%Supplemetal materials

\pagebreak
\widetext
\begin{center}
\textbf{\large Supplemental Materials: Polaron transport and Verwey transition in magnetite}
\end{center}

\setcounter{equation}{0}
\setcounter{figure}{0}
\setcounter{table}{0}
\setcounter{page}{1}
\makeatletter
\renewcommand{\theequation}{S\arabic{equation}}
\renewcommand{\thefigure}{S\arabic{figure}}
\renewcommand{\bibnumfmt}[1]{[S#1]}
\renewcommand{\citenumfont}[1]{S#1}

\section{Low-T kMC calculations}
\begin{figure}[h]
    \centering
    \subfloat{
        \includegraphics[width=0.48\linewidth]{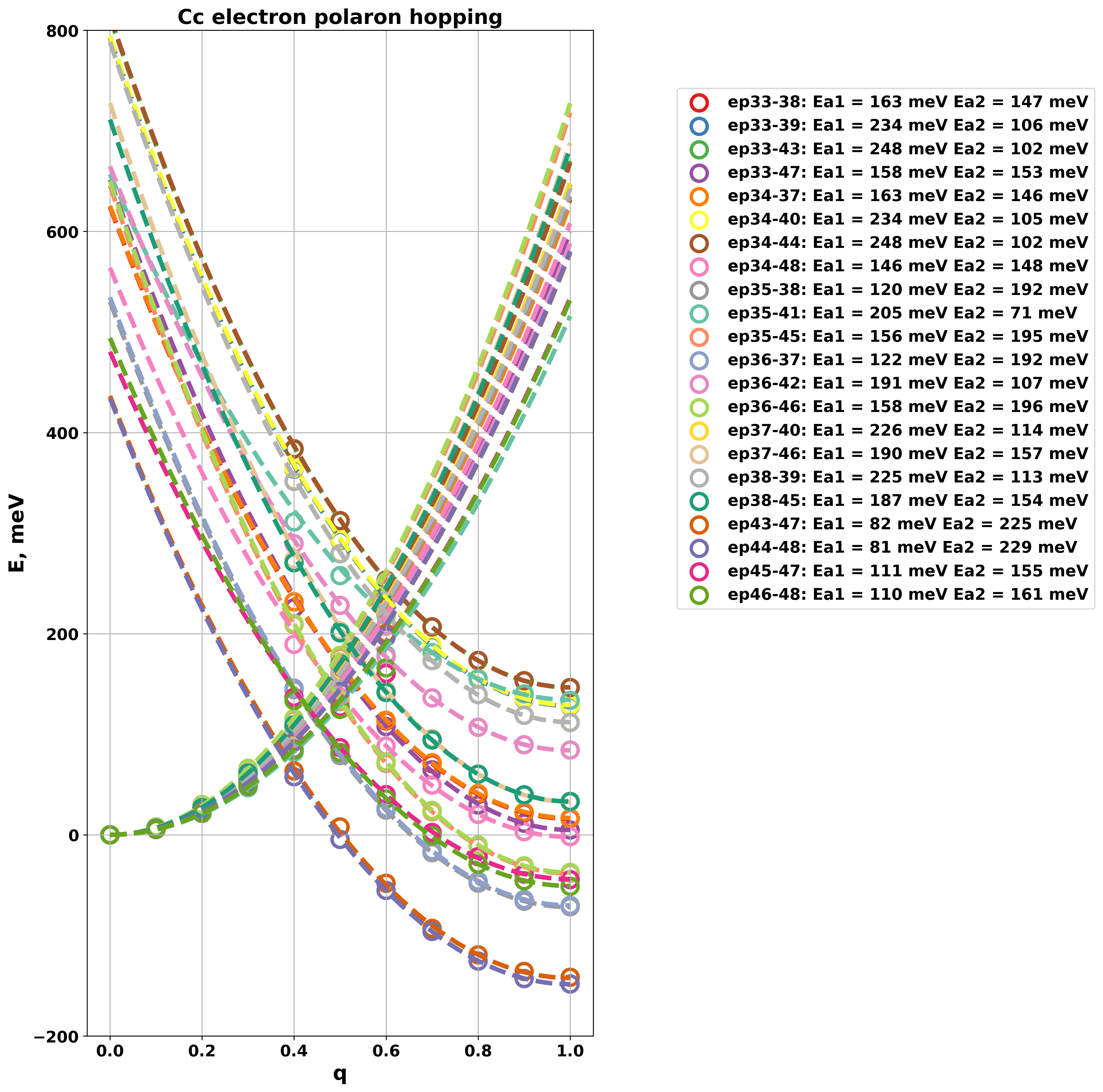}
    }
        \hfill
    \subfloat{
        \includegraphics[width=0.48\linewidth]{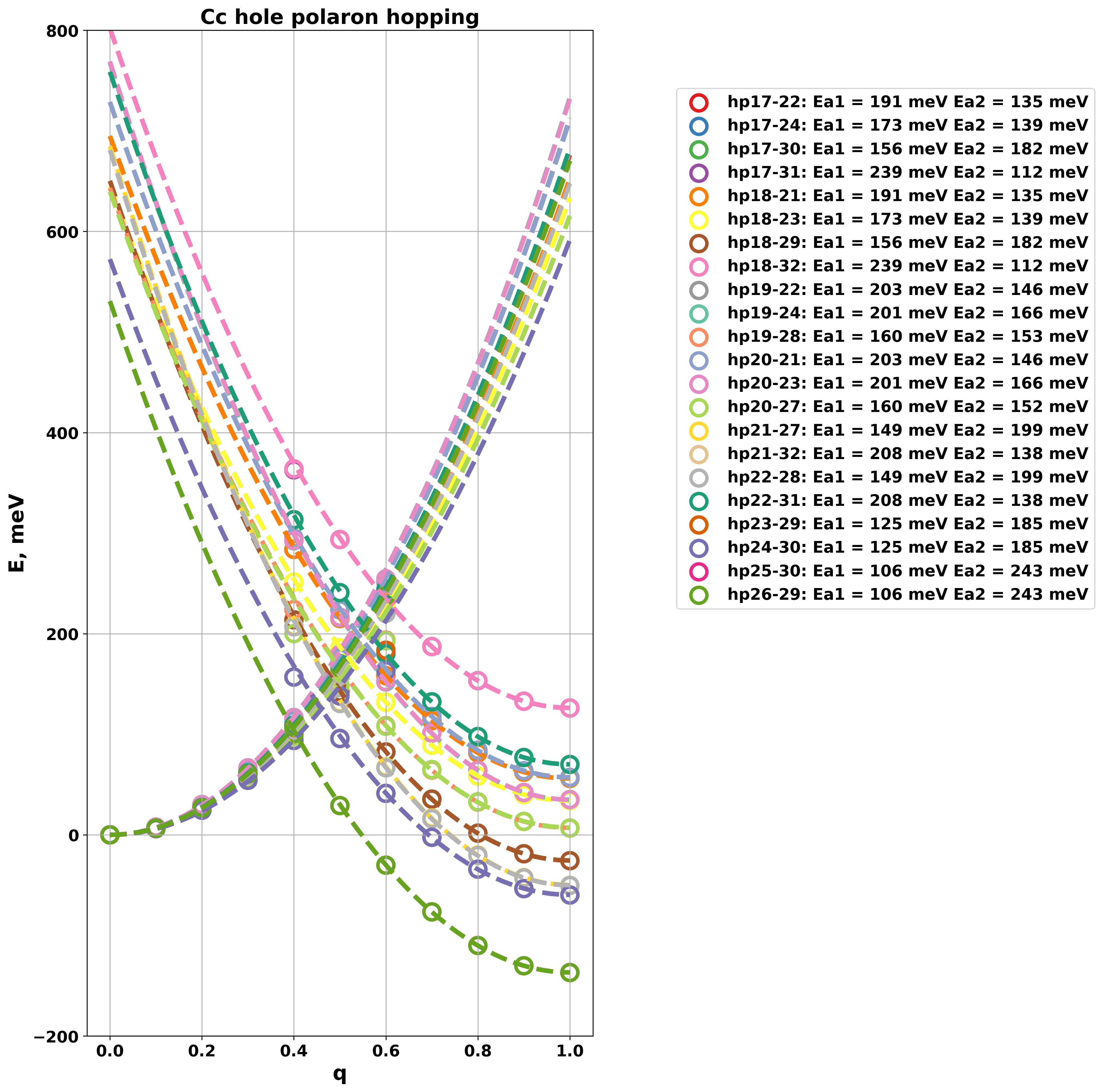}
    }
    \caption{The energy profiles for electron and hole polarons hopping. }
    \label{fig:sccepol}
\end{figure}

This section contains the information needed to reproduce the kMC results. Figure~\ref{fig:sccepol} shows the energy profiles obtained from calculations of static polarons are presented along with the corresponding two activation energy values derived from a parabolic fit (the forward and backward energies differ due to the energy non-equivalence of the sites).

The kMC method is used in a standard rejection-free approach 
For a transition from node $i$ to node $j$, the hopping rate is taken as
\[
k_{ij} = \exp\left(-\frac{E_{a}^{ij}}{k_B T}\right),
\]
where \(E_{a,ij}\) is the activation energy, \(k_B\) is the Boltzmann constant, and \(T\) is the temperature.
For the current node $i$, the total escape rate is
\[
K_i = \sum_{j} k_{ij}.
\]
The probability to select a specific hop $i$ to $j$ is then
\[
P_{ij} = \frac{k_{ij}}{K_i}
= \frac{\exp\left(-E_{a}^{ij}/k_B T\right)}
{\sum\limits_{m} \exp\left(-E_{a}^{im}/k_B T\right)}.
\]
The elapsed time associated with one kMC step is 
\[
\Delta t = -\frac{\ln u}{K_i},
\qquad u \in (0,1),
\]
where $u$ is a uniformly distributed random number.

Thus, temperature-dependent hopping frequencies and corresponding effective activation energies were obtained for electron and hole polarons

\begin{figure}[h!]
    \centering
    \includegraphics[width=0.7\linewidth]{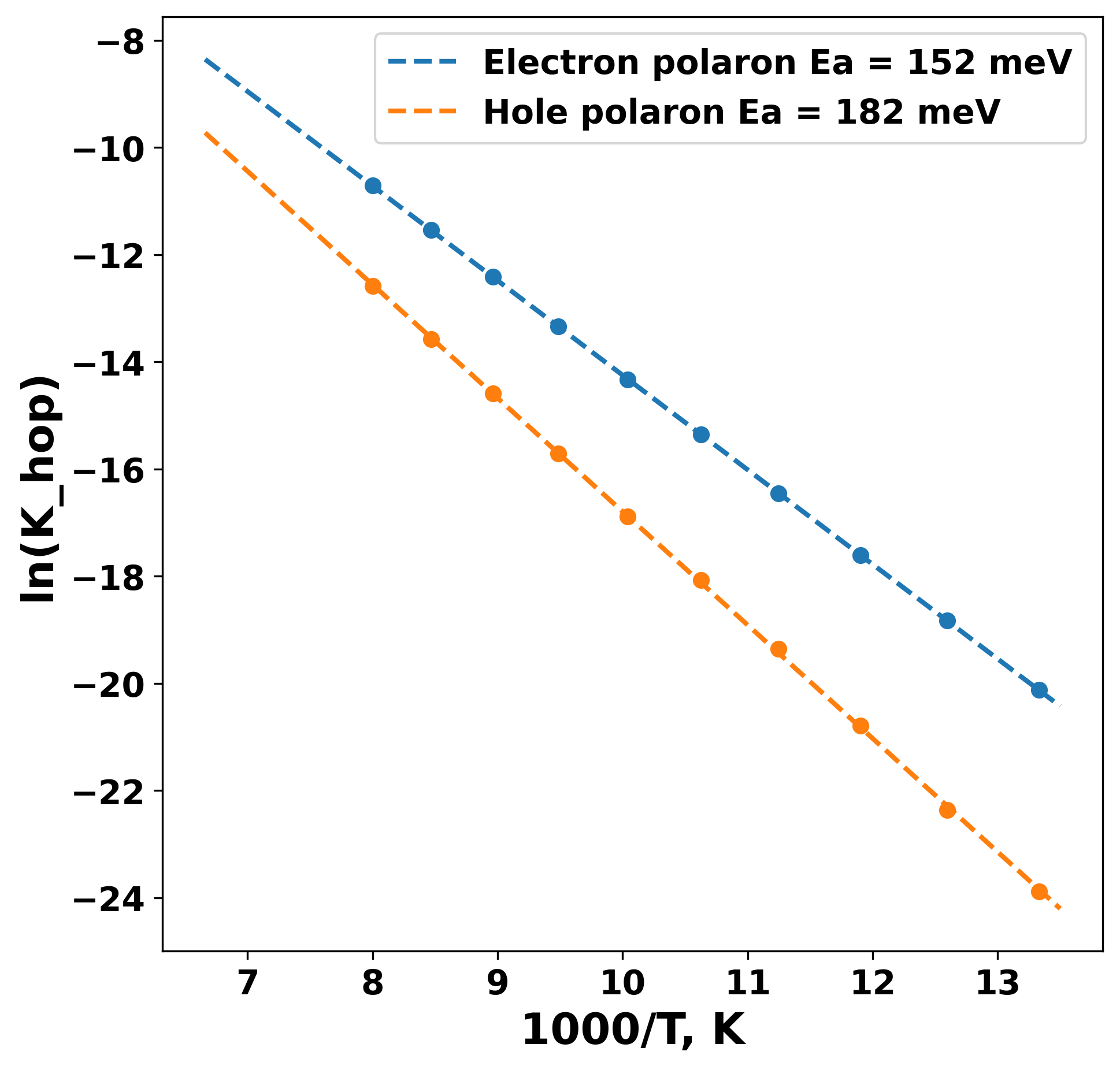}
    \caption{The unnormalized kMC hopping frequencies and corresponding activation energies for electron and hole polarons}
    \label{fig:skmcfreq}  
\end{figure}

\newpage
\section{$U_{eff}$ variance in MD calculations}
\begin{figure}[h]
    \centering
    \subfloat{
        \includegraphics[width=0.29\linewidth]{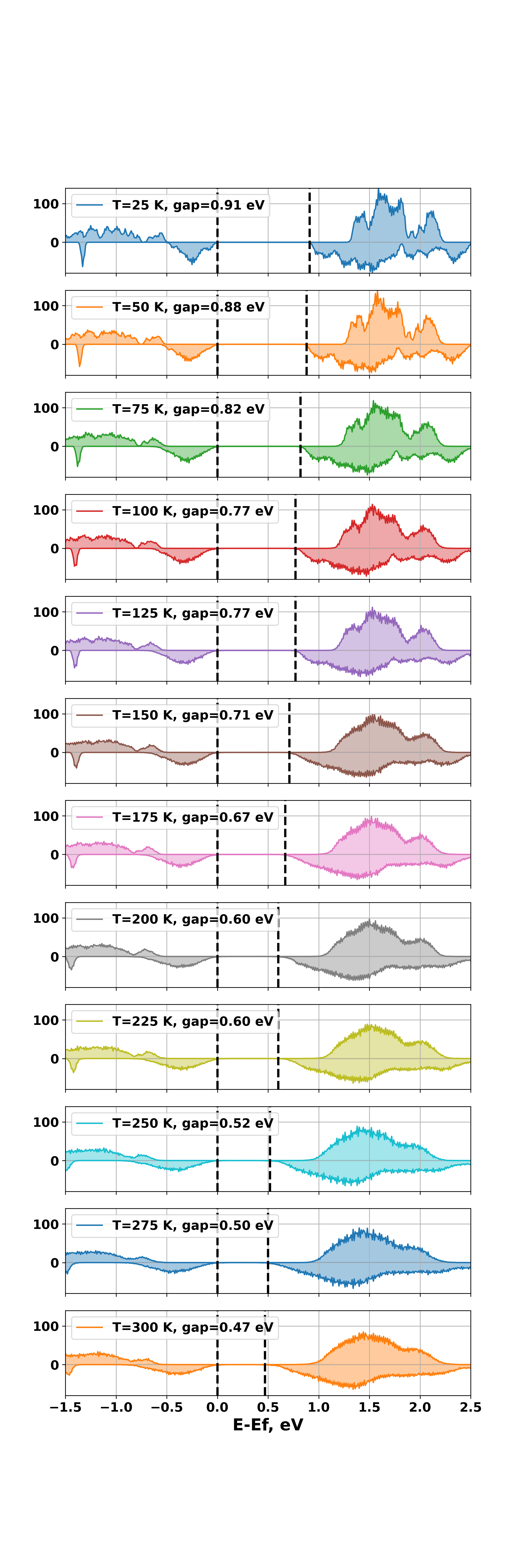}
    }
    \subfloat{
        \includegraphics[width=0.29\linewidth]{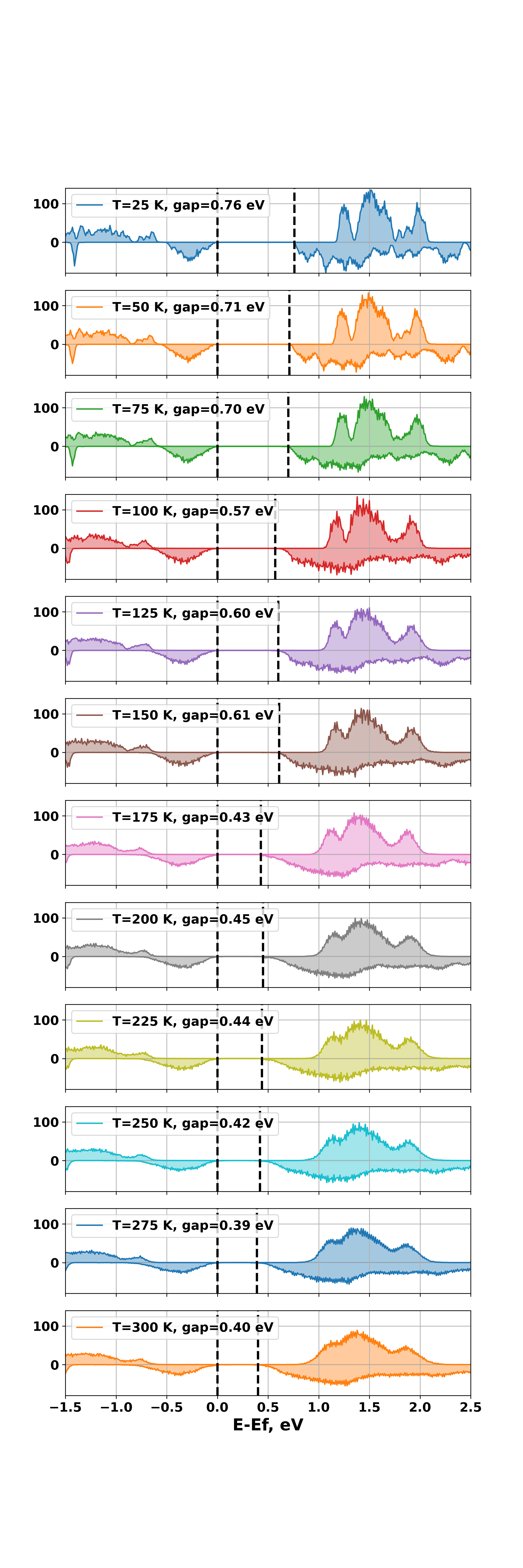}
    }
    \subfloat{
        \includegraphics[width=0.29\linewidth]{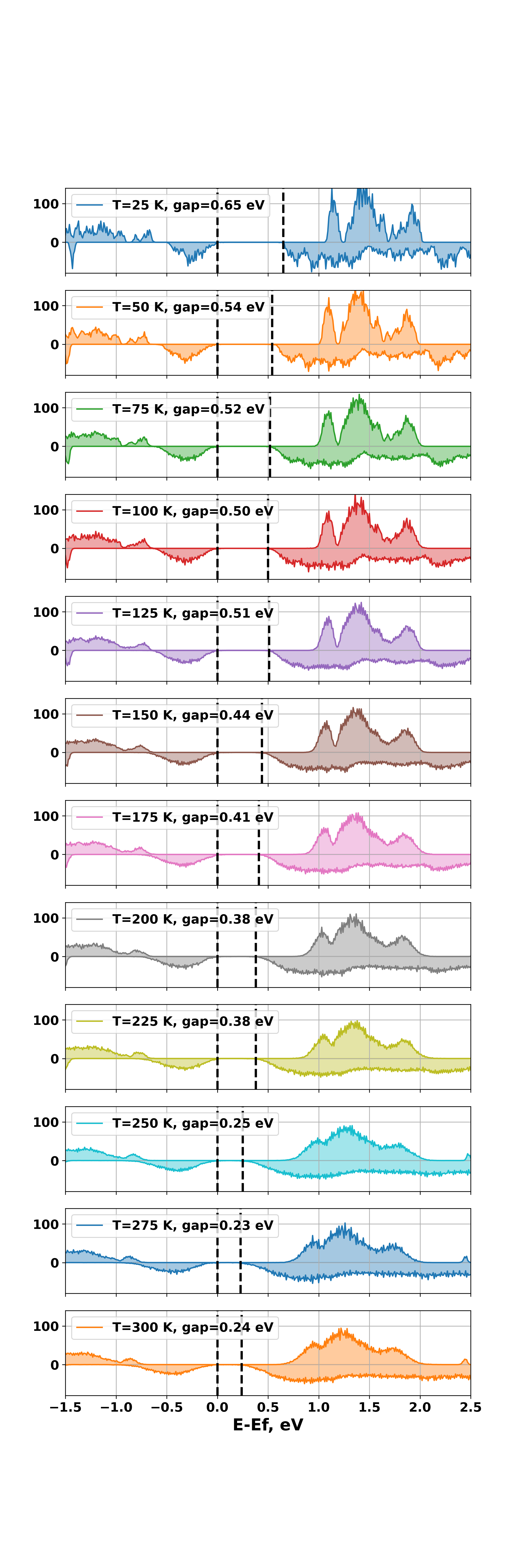}
    }
    \caption{Electronic DOS averaged along MD trajectories for $U_{eff}=3.8, 3.5, 3.2$ eV at given temperature (dashed lines correspond to gap with 1 a.u. threshold) }
    \label{fig:sdos}
\end{figure}

\begin{figure}[h!]
    \centering
    \includegraphics[width=0.7\linewidth]{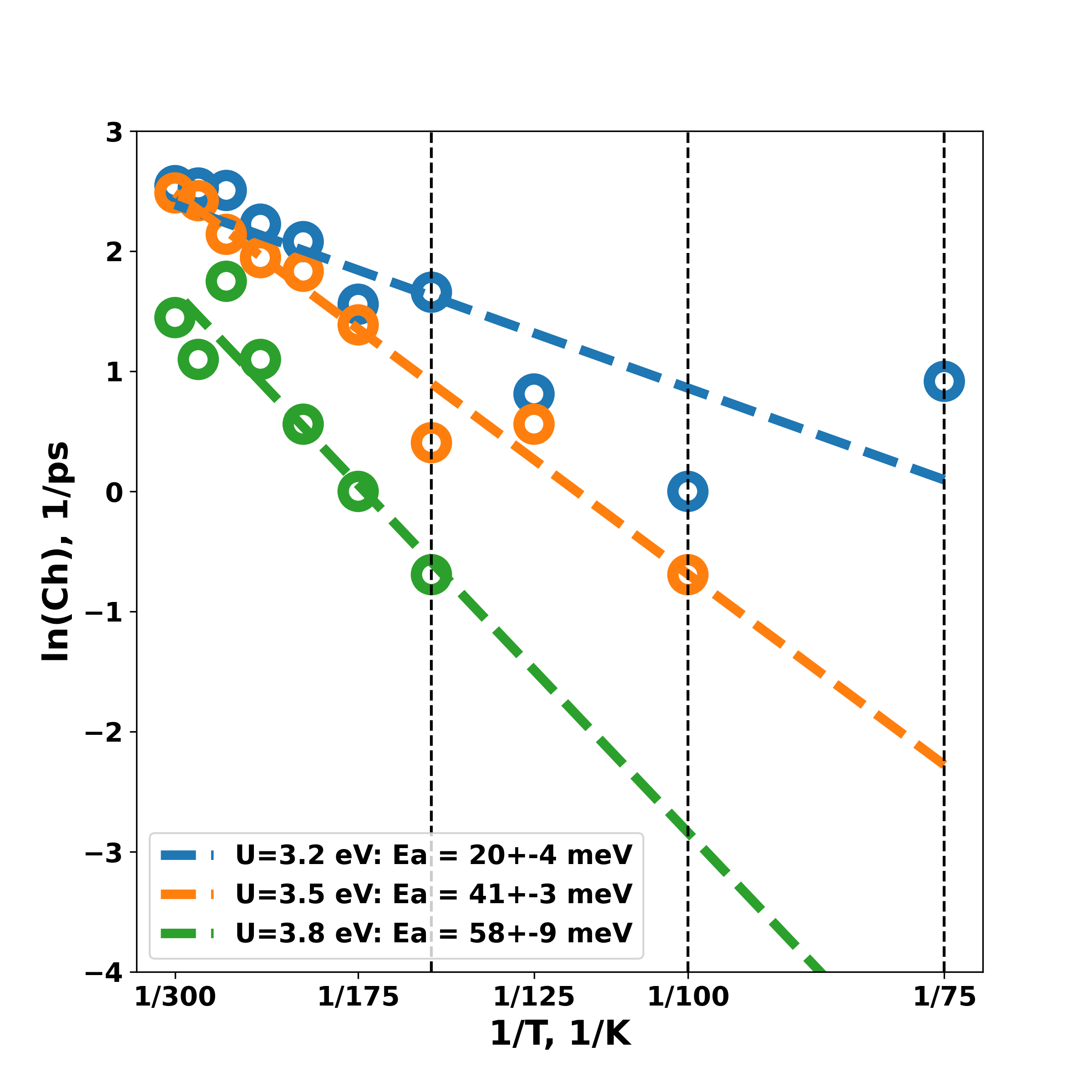}
    \caption{MD trimeron hopping frequencies and corresponding activation energies. Dashed lines indicate the temperature at which the trimeron hopping starts.}
    \label{fig:sarru}  
\end{figure}

This section contains MD calculation results with $U_{eff}$ variation. Electronic DOS data is shown on Figure~\ref{fig:sdos}. Trimeron hopping frequencies are shown on Figure~\ref{fig:sarru}

\end{document}